Category

# A theoretical foundation of the target-decoy search strategy for false discovery rate control in proteomics

Kun He[1,2], Yan Fu[3,*], Wen-Feng Zeng[1,2], Lan Luo[1,2], Hao Chi[1,2], Chao Liu[1,2], Lai-Yun Qing[2], Rui-Xiang Sun[1,2], and Si-Min He[1,2]

[1]Key Lab of Intelligent Information Processing of Chinese Academy of Sciences (CAS), Institute of Computing Technology, CAS, Beijing 100190, China.

[2]University of Chinese Academy of Sciences, Beijing 100049, China.

[3]National Center for Mathematics and Interdisciplinary Sciences, Academy of Mathematics and Systems Science, Chinese Academy of Sciences, Beijing 100190, China.

**ABSTRACT**

**Motivation:** Target-decoy search (TDS) is currently the most popular strategy for estimating and controlling the false discovery rate (FDR) of peptide identifications in mass spectrometry-based shotgun proteomics. While this strategy is very useful in practice and has been intensively studied empirically, its theoretical foundation has not yet been well established.

**Result:** In this work, we systematically analyze the TDS strategy in a rigorous statistical sense. We prove that the commonly used concatenated TDS provides a conservative estimate of the FDR for any given score threshold, but it cannot rigorously control the FDR. We prove that with a slight modification to the commonly used formula for FDR estimation, the peptide-level FDR can be rigorously controlled based on the concatenated TDS. We show that the spectrum-level FDR control is difficult. We verify the theoretical conclusions with real mass spectrometry data.

**Contact:** yfu@amss.ac.cn

## 1 INTRODUCTION

Tandem mass spectrometry (MS/MS) has become the method of choice for identification and quantification of proteins in biological and clinical samples (Aebersold and Mann, 2003; de Godoy, et al., 2008; Weston and Hood, 2004). In this method, after proteolytic digestion of proteins, the resulted peptides are first separated using liquid chromatography, and then are ionized, isolated and fragmented in the mass spectrometer. To identify the peptides, searching the fragmentation mass spectra against a protein sequence database is the standard approach. In this approach, each experimental mass spectrum is computationally compared with the predicted mass spectra of candidate peptides in the database, the peptides are scored and ranked according to their matching degrees to the input spectrum, and the best-scoring peptide is chosen as the identification of the spectrum (Eng, et al., 1994). Usually, a significant proportion of the spectra cannot be correctly identified, because of, for example, the incompleteness of the database searched, the low signal-to-noise ratios of some spectra or simply the imperfect scoring function used by the search engine. Therefore, a critical step of peptide identification is to filter the search results and assess the reliability of the selected identifications (Nesvizhskii, et al., 2007).

The reliability of identifications can be assessed at two different levels, i.e., the single-identification level and the multiple-identification level. The reliability of a single identification is usually assessed using the *p*-value or E-value (Fenyo and Beavis, 2003; Kim, et al., 2008). While for a group of identifications, rather than deciding exactly which identifications are correct or incorrect, a more reasonable approach is to estimate the proportion of incorrect identifications. This is the problem of false discovery rate (FDR) estimation in multiple hypothesis testing (Storey, 2002; Storey, 2003). On the other hand, to generate a list of identifications with their FDR below a given level, e.g. 0.05, the identifications must be filtered using an appropriately determined score threshold. This is the problem of FDR control (Benjamini and Hochberg, 1995).

In the proteomics literature, the FDR is often mistakenly referred to as the false discovery proportion (FDP), i.e., the proportion of incorrect identifications among a set of selected identifications. In fact, FDR is defined as the expectation of FDP in statistics (Benjamini and Hochberg, 1995; Choi and Nesvizhskii, 2008). Although controlling the FDP in each experiment is most desirable, this is in general impossible (Benjamini and Hochberg, 1995). However, controlling the FDR is more feasible. Moreover, FDR estimation and FDR control are also two different concepts that are rigorously defined (Storey, 2002; Storey, et al., 2004). FDR estimation is to calculate an FDR for a given score threshold. Specially, if the expectation of the estimated FDR is not less than the real FDR, we say the estimate is conservative. FDR control is to search for a score threshold such that the real FDR does not exceed the given FDR level.

At present, the target-decoy search (TDS) strategy is the gold standard for estimating and controlling the FDR of peptide identifications in proteomics (Elias and Gygi, 2007; Walzthoeni, et al., 2012; Wilhelm, et al., 2014). In this strategy (Fig. 1), in addition to the original protein database (target database), the experimental mass spectra are also searched against an equal-size decoy data-base, which is usually constructed by reversing or shuffling the amino acid sequences in the target database. Because the protein sequences in the decoy database cannot be present in the sample, any matches to these sequences must be incorrect identifications, and can be used to estimate the number of incorrect matches to target sequences.

*To whom correspondence should be addressed.

There are two different modes to perform a TDS, i.e., the concatenated TDS and the separate TDS (Kall, et al., 2008). The concatenated TDS is performed by searching a combined database of the target and decoy sequences. With this search mode, only one identification will be produced for each spectrum (the best-scoring match to either the target or the decoy sequences). Consider the identifications scoring equal to or better than $x$. Let $N_{tar}(x)$ be the total number of identifications from the target sequences (called target identifications), $N_{dec}(x)$ be the number of identifications from the decoy sequences (called decoy identifications), and $N_{inc}(x)$ be the number of incorrect target identifications. Thus, the FDR for a fixed score threshold $x$, $FDR(x)$, is $E[N_{inc}(x)/max\{N_{tar}(x), 1\}]$. $N_{inc}(x)$ is of interest but is unknown. Usually, because there is an equal chance for an incorrect identification to be a target or a decoy identification (Equal Chance Assumption) (Elias and Gygi, 2007; Elias and Gygi, 2010), one can estimate $N_{inc}(x)$ with $N_{dec}(x)$, and further estimate the FDR of target identifications scoring equal to or better than $x$ as

$$N_{dec}(x)/N_{tar}(x). \quad (1)$$

The separate TDS is performed by searching the target database and the decoy database separately. With this search mode, two identifications will be produced for each spectrum. One of them is the target identification (the best-scoring match to the target sequences) and the other is the decoy identification (the best-scoring match to the decoy sequences). The FDR at the score threshold $x$ is estimated as $\pi_0 N_{dec}(x)/N_{tar}(x)$, where $\pi_0$ is the proportion of incorrect ones among target identifications (Kall, et al., 2008). Compared to the concatenated TDS, the separate TDS is less widely used. In this article, we will concentrate on the concatenated TDS.

In MS/MS, one peptide may be fragmented and detected multiple times, resulting in many similar redundant mass spectra. Consequently, one peptide may be identified multiple times by redundant spectra, even if the identification is incorrect (Nesvizhskii, et al., 2003). Thus the FDRs of identifications at the peptide level and the spectrum level are different and should be estimated differently. When estimating the spectrum-level FDR, the identifications from all spectra are retained. When estimating the peptide-level FDR, the identifications of the same peptide are counted only once. This is usually done by retaining the best-scoring identification of a peptide and weeding out other redundant identifications (Cox and Mann, 2008; Granholm, et al., 2013). After abundance removal, any two incorrectly identified spectra must be matched to different peptides, and whether one spectrum is matched to the target sequences is independent of the other (Independence Assumption). In this article, the FDR of peptide identifications is used to refer to both the peptide-level FDR and the spectrum-level FDR.

In addition to FDR estimation, the TDS strategy is also widely used for FDR control. Given a FDR control level $\alpha$, a commonly used procedure to determine the score threshold is as follows. Without loss of generality, assume that the scoring function is defined such that larger scores are better. After performing a concatenated TDS, FDR estimation is carried out at varying scoring thresholds, and the lowest $x$ at which the estimated FDR is no more than $\alpha$ is chosen as the final score threshold $t$. All target identifications scoring no less than $t$ are retained while other identifications are filtered out. In formal words, with the concatenated TDS, the score threshold is determined as $t = min\{x | N_{dec}(x)/N_{tar}(x) \leq \alpha\}$. If there is no such $x$, $t$ is set as larger than the highest score of all identifications and then all identifications are filtered out.

While the TDS strategy is very popular and useful in practice and has been intensively studied empirically (Elias and Gygi, 2007; Granholm, et al., 2013; Jeong, et al., 2012), its theoretical foundation has not yet been well established. For example, it is unknown whether the estimate made by the TDS is conservative or optimistic, or whether the FDR can be rigorously controlled based on the TDS. Note that rigorous FDR control is very important and desirable from the statistical point of view. Moreover, the FDR can be estimated and controlled either at the spectrum level or at the peptide level. Which level the FDR should be controlled at is still unsettled. Without clear answers to these questions, proteomics would still be a less-strict science.

In this work, we systematically analyze the TDS strategy in a rigorous statistical sense. We prove that the common way to use the concatenated TDS provides a conservative estimate of the FDR, but it cannot rigorously control the FDR. We prove that with a slight modification to the commonly used formula for FDR estimation, the peptide-level FDR can be rigorously controlled based on the concatenated TDS. Furthermore, we show that the spectrum-level FDR control is difficult. We verify the theoretical conclusions with real MS/MS data. In summary, our work provides a strict theoretical foundation for the concatenated TDS and thoroughly answers some key questions pending in the field.

## 2 THEORETICAL ANALYSIS

### 2.1 FDR of peptide identifications

FDR is defined as the expectation of the FDP. Here we first clarify the probability space for calculating the FDR of peptide identifications.

When performing the database search, a spectrum is usually compared with many candidate peptides and each peptide makes up a peptide-spectrum match with the spectrum. For each spectrum, either all matches are incorrect, or only one of them is correct. Only the best-scoring peptide-spectrum match is chosen as the identification of the spectrum. In calculating the FDR, the scores of incorrect matches can be regarded as random variables. Then the scores of incorrect identifications are also random variables. At the same time, even for a spectrum that has a correct match, there may be one or more incorrect matches whose scores are larger than the correct one (in this case the spectrum is incorrectly identified). Namely, with some probability a spectrum with a correct match will be incorrectly identified. In this context, for a given score threshold $x$, the number of incorrect identifications scoring equal to or better than $x$ is a random variable, so is the total number of identifications. Thus, the proportion of incorrect identifications scoring equal to or better than $x$, $FDP(x)$, is also a random variable. The FDR for score $x$, $FDR(x)$, is the expectation of $FDP(x)$. If the expectation of the estimated FDR is not less than the real FDR, i.e., $E[\widehat{FDR}(x)] \geq FDR(x)$, then we say $\widehat{FDR}(x)$ is a conservative estimate of $FDR(x)$. Because $FDR(x) = E[FDP(x)]$, we have that, on average, a conservative estimate of the FDR does not underestimate the proportion of incorrect identifications.

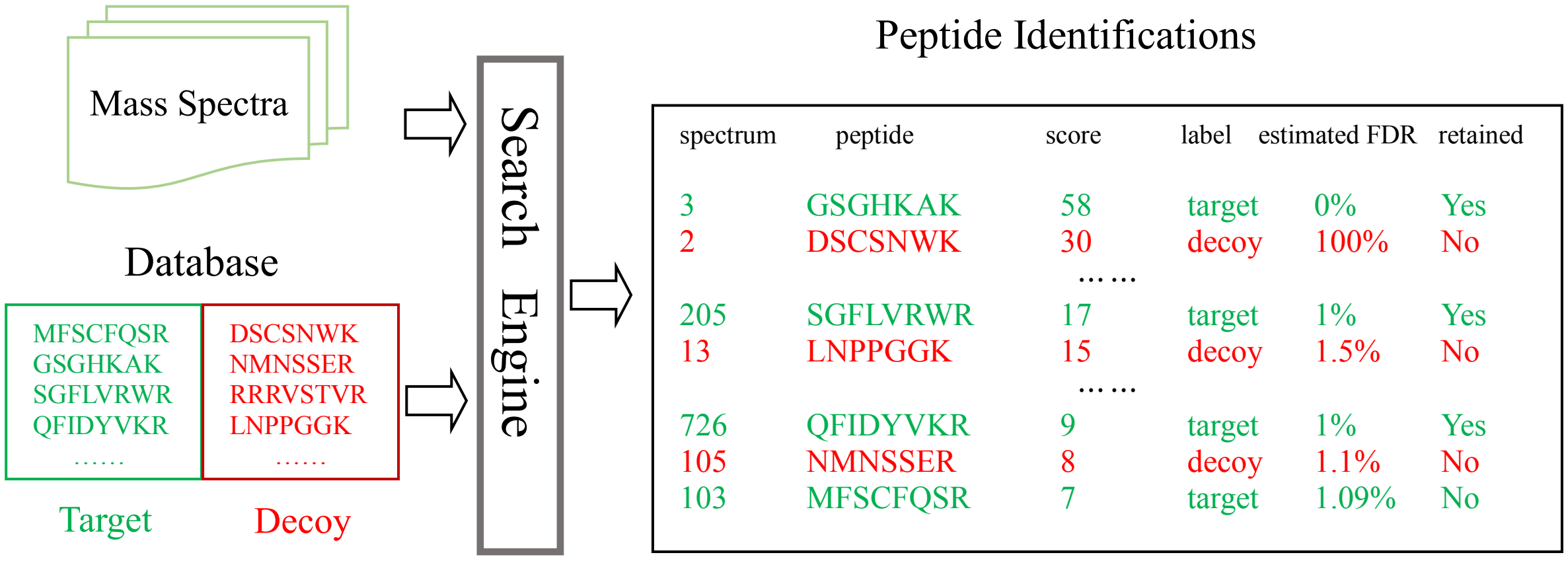

**Figure 1.** A workflow of the TDS strategy for FDR estimation and FDR control. For the concatenated TDS, the FDR of target identifications scoring equal to or better than $x$ is estimated as $N_{dec}(x)/N_{tar}(x)$. To control the FDR at $\alpha$, the lowest $x$ at which the estimated FDR is no more than $\alpha$ is chosen as the final score threshold. In the illustrated example, the score threshold for the 1% FDR control level is 9, although the estimated FDRs for score 30 and score 15 are both larger than 1%.

Similarly, for a given FDR level $\alpha$, let $t$ be the score threshold determined according to some peptide identification filtering criterion. The FDP of identifications scoring equal to or better than $t$, is also a random variable, and the FDR is the expectation of FDP. An FDR control method should be able to choose a reasonable $t$ such that $E(FDP) = FDR < \alpha$. That is, on average, the proportion of incorrect identifications scoring equal to or better than the threshold does not exceed the given FDR level.

For the concatenated TDS, there is an equal chance for an incorrect identification to be a target or a decoy identification. Thus, $N_{inc}(x)$, $N_{dec}(x)$ and $N_{tar}(x)$ are all random variables and $N_{dec}(x)/N_{tar}(x)$ is also a random variable. The expectation of the estimated FDR for $x$ is $E[N_{dec}(x)/N_{tar}(x)]$ and the real FDR for $x$ is $E[N_{inc}(x)/N_{tar}(x)]$. To determine whether the concatenated TDS provides a conservative estimate of the FDR is to determine whether $E[N_{dec}(x)/N_{tar}(x)]$ is no less than $E[N_{inc}(x)/N_{tar}(x)]$. Note that $N_{tar}(x)$ can be zero. The discussion for this case is provided in Supplementary information.

Similarly, assume that the score threshold determined with the TDS for FDR control level $\alpha$ is $t$, the real FDR is $E[N_{inc}(t)/max\{N_{tar}(t), 1\}]$. To determine whether the concatenated TDS controls the FDR is to determine whether $E[N_{inc}(t)/max\{N_{tar}(t), 1\}]$ is no more than $\alpha$.

## 2.2 FDR estimation

Though the concatenated TDS has been widely used to estimate the FDR of peptide identifications, it is still unknown whether the estimate made by the TDS is conservative or optimistic. The following theorem shows that the concatenated TDS provides a conservative estimate of the FDR. Namely, on average, the concatenated TDS does not underestimate the proportion of incorrect identifications. To some extent, this explains the effectiveness of TDS in practice.

*Theorem 1 (FDR Estimation Theorem). Under the Equal Chance Assumption, $N_{dec}(x)/N_{tar}(x)$ is a conservative estimate of the FDR of peptide identifications scoring equal to or better than $x$.*

The formal proof of Theorem 1 is given in Supplementary information, sections 1. Here we give a simplified proof, in which we assume that the Independence Assumption holds and $N_{tar}(x)$ is larger than 0.

Consider the identifications scoring equal to or better than $x$. Let $N_{cor}(x)$ be the number of correct target identifications, and $P_x(a,b,c) = P[N_{inc}(x) = a, N_{dec}(x) = b, N_{cor}(x) = c]$, where $a$, $b$ and $c$ are nonnegative integers. If we estimate the FDR with Eq. (1), the expectation of the estimated FDR is

$$E[\widehat{FDR}(x)] = \sum_{a,b,c} \frac{b}{a+c} P_x(a,b,c). \quad (2)$$

According to the Equal Chance Assumption and the Independence Assumption, for any case where $N_{inc}(x) = a, N_{dec}(x) = b, N_{cor}(x) = c$ and $a < b$, there is a symmetric case where $N_{inc}(x) = b, N_{dec}(x) = a, N_{cor}(x) = c$, and the probabilities of these two symmetric cases are equal, i.e., $P_x(a,b,c) = P_x(b,a,c)$. Therefore,

$$\widehat{FDR}(x) = \sum_{a<b,c} \frac{b}{a+c} P_x(a,b,c) + \frac{a}{b+c} P_x(b,a,c) + \sum_{a=b,c} \frac{b}{a+c} P_x(a,b,c)$$
$$= \sum_{a<b,c} \left(\frac{b}{a+c} + \frac{a}{b+c}\right) P_x(a,b,c) + \sum_{a=b,c} \frac{b}{a+c} P_x(a,b,c). \quad (3)$$

On the other hand, the real FDR is

$$FDR(x) = \sum_{a,b,c} \frac{a}{a+c} P_x(a,b,c). \quad (4)$$

Similar to the expectation of the estimated FDR, we have

$$FDR(x) = \sum_{a<b,c}\left(\frac{a}{a+c}+\frac{b}{b+c}\right)P_x(a,b,c)+\sum_{a=b,c}\frac{a}{a+c}P_x(a,b,c). \quad (5)$$

Because

$$\frac{b}{a+c}+\frac{a}{b+c}=\frac{a^2+b^2+ac+bc}{(a+c)(b+c)} \geq \frac{2ab+ac+bc}{(a+c)(b+c)}=\frac{a}{a+c}+\frac{b}{b+c}, \quad (6)$$

we have

$$E\left[\widehat{FDR(x)}\right] = \sum_{a<b,c}\left(\frac{b}{a+c}+\frac{a}{b+c}\right)P_x(a,b,c)+\sum_{a=b,c}\frac{b}{a+c}P_x(a,b,c) \geq \sum_{a<b,c}\left(\frac{a}{a+c}+\frac{b}{b+c}\right)P_x(a,b,c)+\sum_{a=b,c}\frac{b}{a+c}P_x(a,b,c) = \sum_{a<b,c}\left(\frac{a}{a+c}+\frac{b}{b+c}\right)P_x(a,b,c)+\sum_{a=b,c}\frac{a}{a+c}P_x(a,b,c) = FDR(x). \blacksquare \quad (7)$$

Namely, the real FDR is no greater than the expectation of the estimated FDR for fixed $x$. Thus, Eq. (1) is a conservative estimate of the FDR.

## 2.3 FDR Control

It is known that some conservative estimates of the FDR can be used to define valid FDR control procedures (Storey, et al., 2004). In proteomics, the conservative estimate given in Eq. (1) has been indeed used for FDR control. That is, peptide identifications are filtered by searching for the lowest score threshold such that the estimated FDR with Eq. (1) is not greater than the specified FDR level. A crucial question is whether this filtering criterion based on Eq. (1) can really control the FDR. There was no answer in the past. Actually, we can prove the following proposition.

*Proposition 1. For a given FDR level $\alpha$, filtering out all target identifications scoring worse than $t = min\{x|N_{dec}(x)/N_{tar}(x) \leq \alpha\}$ cannot control the FDR.*

Proof. We prove the proposition with a counter-example. Suppose that we have *n* identifications after performing a concatenated TDS on a data set and the user-specified FDR control level is 1%. Suppose that all the identifications are incorrect for some reason, for example, a wrong database used. Let $y$ be the score of the best-scoring identification. Because all identifications are incorrect, according to the Equal Chance Assumption, the probability of the best-scoring identification being a target identification is 0.5. When this happens, we have $N_{dec}(y) = 0$, $N_{tar}(y) = 1$ and the estimated FDR for score $y$ is $N_{dec}(y)/N_{tar}(y) = 0/1 = 0 < 1\%$. Then the final score threshold is no more than $y$, and at least the best-scoring identification will be retained according to the filtering criterion defined in Proposition 1. In this case, the FDP is 100%, since all identifications are incorrect. On the other hand, the probability of the best-scoring identification being a decoy identification is also 0.5. When this happens, the FDP is not less than 0. Therefore, the expected FDP, or FDR, is not less than $0.5 \times 1 + 0.5 \times 0 = 50\%$, far higher than the FDR control level (1%). ■

As shown in Proposition 1, Eq. (1) cannot control the FDR, either at the spectrum level or the peptide level. Though it is possible to control the FDR based on a conservative estimate, some modifications may be needed to make the FDR control strategy stricter (Storey, et al., 2004). In fact, by modifying Eq. (1) into

$$[N_{dec}(x)+1]/N_{tar}(x), \quad (8)$$

we obtain a more conservative estimate of the FDR. Based on the new estimate, we can define a filtering criterion for the peptide-level FDR control.

*Theorem 2 (FDR Control Theorem). Under the Equal Chance Assumption and the Independence Assumption, for any given FDR level $\alpha$, filtering out all target identifications scoring worse than $t = min\{x|[N_{dec}(x)+1]/N_{tar}(x) \leq \alpha\}$ controls the peptide-level FDR.*

The formal proof of Theorem 2 is given in Supplementary information, sections 1. Here we give an equivalent less formal proof. Because the identifications have been sorted by their scores, we refer to the *i*-th best-scoring identification as the *i*-th identification. Let $T_1$ be the number of incorrect target identifications scoring better than the best-scoring decoy identification. If there are no decoy identifications, $T_1$ is the number of all incorrect target identifications. Assume that there are $N$ incorrect identifications in total, including incorrect target identifications and decoy identifications. If for any constant $n$,

$$E\left[\frac{N_{inc}(t)}{max\{N_{tar}(t),1\}}|N=n\right] < \alpha, \quad (9)$$

from the law of total expectation, we have

$$E\left[\frac{N_{inc}(t)}{max\{N_{tar}(t),1\}}\right] = E\left\{E\left[\frac{N_{inc}(t)}{max\{N_{tar}(t),1\}}|N=n\right]\right\} < \alpha \quad (10)$$

Therefore, we only need to prove the theorem in the case that the number of incorrect identifications is constant $n$.

Firstly, we prove that $E(T_1) < 1$. If the Equal Chance Assumption and the Independence Assumption hold, for any $0 < i \leq n$, the probability that the $i-1$ best-scoring incorrect identifications are target identifications and the $i$-th one is a decoy identification is $0.5^{i-1} \times 0.5 = 0.5^i$. Or formally, $P(T_1 = i-1) = 0.5^i$. The probability that all incorrect identifications are target identifications is $0.5^n$. Or formally, $P(T_1 = n) = 0.5^n$. Then we have

$$E(T_1) = 0.5^n n + \sum_{i=1}^{n} 0.5^i(i-1) = 1 - 0.5^n < 1. \quad (11)$$

Secondly, we prove that $E(T_1) \geq E\{N_{inc}(t)/[N_{dec}(t)+1]\}$. There are three different cases: $N_{inc}(t) = N_{dec}(t) = 0$, $N_{inc}(t) + N_{dec}(t) = n$ and $0 < N_{inc}(t) + N_{dec}(t) < n$. If $N_{inc}(t) = N_{dec}(t) = 0$, we have

$$E[T_1|N_{inc}(t) = N_{dec}(t) = 0] \geq 0. \quad (12)$$

If $N_{inc}(t) + N_{dec}(t) = n$, all identifications pass the score threshold and we have

$$E[T_1|N_{inc}(t) = a, N_{dec}(t) = b, a+b=n] = \frac{a}{b+1}. \quad (13)$$

This is because the $b$ decoy identifications divide the $a$ incorrect target identifications into $b+1$ groups according to their scores, and the expected numbers of incorrect target identifications in these groups are the same, all equal to $a/(b+1)$. If $0 < N_{inc}(t) + N_{dec}(t) < n$, at least one identification passes the score threshold. According to the definition of $t$, we have $[N_{dec}(t)+1]/N_{tar}(t) \leq \alpha$. Note that the best-scoring identification that scores worse than the score threshold must be a decoy identification. Otherwise, if it is a target identification with a score $y$, then $y < t$ and

$$\frac{N_{dec}(y)+1}{N_{tar}(y)} = \frac{N_{dec}(t)+1}{N_{tar}(t)+1} < \frac{N_{dec}(t)+1}{N_{tar}(t)} \leq \alpha. \quad (14)$$

This is contradictive to $t = min\{x|[N_{dec}(x)+1]/N_{tar}(x) \le \alpha\}$. In other words, there are $N_{inc}(t)$ incorrect target identifications scoring better than the $[N_{dec}(t)+1]$-th decoy identification. Therefore, it can be proved that

$$E[T_1|N_{inc}(t)=a, N_{dec}(t)=b, 0<a+b<n] = \frac{a}{b+1}. \quad (15)$$

Then from Eq. (12), Eq. (13) and Eq. (15), we have that no matter what integer values $a$ and $b$ take,

$$E[T_1|N_{inc}(t)=a, N_{dec}(t)=b] \ge \frac{a}{b+1}. \quad (16)$$

Therefore, from the law of total expectation, we have

$$E(T_1) = E\{E[T_1|N_{inc}(t)=a, N_{dec}(t)=b]\}$$
$$\ge \sum_{a,b} \frac{a}{b+1} P[N_{inc}(t)=a, N_{dec}(t)=b] = E\left[\frac{N_{inc}(t)}{N_{dec}(t)+1}\right]. \quad (17)$$

Thirdly, from Eq. (11) and Eq. (17), we have

$$E\left[\frac{N_{inc}(t)}{N_{dec}(t)+1}\right] \le E(T_1) < 1. \quad (18)$$

Because the score threshold is defined as $t = min\{x|[N_{dec}(x)+1]/N_{tar}(x) \le \alpha\}$, if any target identification passes the score threshold, we have

$$N_{tar}(t) \ge [N_{dec}(t)+1]/\alpha. \quad (19)$$

Then

$$\frac{N_{inc}(t)}{N_{tar}(t)} \le \frac{\alpha N_{inc}(t)}{N_{dec}(t)+1}. \quad (20)$$

If no target identification passes the score threshold, then $N_{inc}(t) = 0$ and it can also be proved that

$$\frac{N_{inc}(t)}{max\{N_{tar}(t),1\}} = 0 = \frac{\alpha N_{inc}(t)}{N_{dec}(t)+1}. \quad (21)$$

Therefore, whether there are some target identifications passing the score threshold or not, we always have

$$\frac{N_{inc}(t)}{max\{N_{tar}(t),1\}} \le \frac{\alpha N_{inc}(t)}{N_{dec}(t)+1}. \quad (22)$$

Thus,

$$E\left[\frac{N_{inc}(t)}{max\{N_{tar}(t),1\}}\right] \le \alpha E\left[\frac{N_{inc}(t)}{N_{dec}(t)+1}\right]. \quad (23)$$

From Eq. (18), we have

$$E\left[\frac{N_{inc}(t)}{max\{N_{tar}(t),1\}}\right] \le \alpha E\left[\frac{N_{inc}(t)}{N_{dec}(t)+1}\right] \le \alpha E(T_1) < \alpha. \quad (24)$$

The FDR is controlled.■

The filtering criterion based on Eq. (8) can control the peptide-level FDR, but it cannot control the spectrum-level FDR. This can be demonstrated by a counter-example. Suppose $n$ ($n$ is larger than 20) spectra generated from one peptide are analyzed with the concatenated TDS. The user-specified spectrum-level FDR control level is 5%. After the search, we obtain $n$ identifications. Let $y$ be the lowest score of these identifications. Suppose that all the identifications are incorrect for some reason, for example, a wrong database used. In addition, because the spectra generated by the same peptide are very similar to each other, we can assume that they are matched to the same peptide. As a result, the probabilities of this identified peptide coming from the target sequences and the decoy sequences are the same (0.5). If it is from the target sequences, we have $N_{dec}(y) = 0$, $N_{tar}(y) = n$ and the estimated FDR for score $y$ is $[N_{dec}(y)+1]/N_{tar}(y) = (0+1)/n < 5\%$. Then the final score threshold is no more than $y$, and all of the $n$ identifications will be retained. In this case, the FDP is 100%. If the peptide is from the decoy sequences, the FDP is not less than 0. Thus, the FDR is not less than $0.5 \times 1 + 0.5 \times 0 = 50\%$. Therefore, the spectrum-level FDR is not controlled. The identifications from redundant spectra are correlated, even if they are incorrect identifications (Granholm, et al., 2013). These correlations make the spectrum-level FDR control very difficult if not impossible.

Consider the following FDR estimator

$$[N_{dec}(x)+c]/N_{tar}(x). \quad (25)$$

Obviously, for any $c > 0$, Eq. (25) is a conservative estimate of the FDR. However, only when $c \ge 1$, the Eq. (25)-based criterion can control the FDR.

*Theorem 3 (Optimal FDR Control Theorem). For any* $c < 1$, *filtering out all target identifications scoring worse than* $t = min\{x|\ [N_{dec}(x)+c]/N_{tar}(x) \le \alpha\}$ *cannot control the FDR.*

Moreover, for any $c > 1$, Eq. (25) is larger than Eq. (8). Therefore, with Eq. (25)-based criterion, the number of correct identifications we can obtain is no more than that we obtain with the Eq. (8)-based criterion. That is to say, the Eq. (8)-based criterion is optimal to some extent. (See Supplementary information, sections 1 for the proof of Theorem 3.)

## 3 EXPERIMENTAL VALIDATION

Two data sets were used to verify the theoretical conclusions above. One was the Worm data set, and the other one was the Human data set. The Worm data set contains 13,545 MS/MS spectral pairs derived from a whole cell lysate of C. elegans on an ETD enabled LTQ-Orbitrap XL mass spectrometer (Chi, et al., 2013; Chi, et al., 2010). Each spectral pair contains an HCD spectrum and an ETD spectrum from the same precursor ion. A concatenated TDS was performed with pFind 2.8 (Fu, et al., 2004; Wang, et al., 2007). The database was downloaded from ftp://ftp.wormbase.org/pub/wormbase/releases/WS231/species/c_elegans/ and the decoy database is constructed by protein sequence reversal. The score thresholds corresponding to the 1% estimated peptide-level FDR were calculated for the HCD spectra and the ETD spectra separately. Only the identifications passing the score thresholds were retained. We selected identification pairs corresponding to the same peptides and obtained 4,568 reliable HCD identifications. In the following, we will refer to these 4,568 HCD spectra as reliably identified spectra (RI spectra), and the remaining 8977 HCD spectra as unreliably identified spectra (UI spectra).

The Human data set contains 7,833 MS/MS HCD spectra derived from HEK293 cells on an Orbitrap Velos mass spectrometer (Frese, et al., 2011). Concatenated TDSs were performed with pFind 2.8 and MaxQuant 1.3.0.5 (Cox and Mann, 2008). The database was downloaded from Uniprot on July 11th, 2013 and the decoy database is constructed by protein sequence reversal. The score thresholds corresponding to the 1% estimated peptide-level FDR for pFind and MaxQuant were calculated, respectively. Only the identifications passing the corresponding thresholds were retained. Identical identifications from pFind and MaxQuant were selected, resulting in a set of 4,879 reliable identifications. The remaining 2954 spectra constituted the set of UI spectra.

The spectra were sampled and searched repeatedly so that the real FDR could be computed from the averaged FDP. Our purpose was to evaluate different filtering criteria by comparing the real FDRs controlled with them to the specified FDR control level (5% in this experiment). For a filtering criterion, the FDRs are calculated for different spectra sizes. The detailed process of calculating an FDR for $n$ RI spectra is as follows. First, $n$ RI spectra and enough UI spectra were sampled. The UI spectra were altered by shifting their precursor ion m/z values by some mass deviation, for example, 10 Da, rendering them impossible to be correctly identified by database search (Elias and Gygi, 2007). Second, the sampled RI spectra and UI spectra were merged together and searched with pFind 2.8. The RI spectra were correctly identified again while the UI spectra were incorrectly identified. The identifications were filtered at 5% FDR control level. Because we knew which of the retained identifications were correct, we could calculate the FDP. Third, by re-sampling the UI spectra and shifting their precursor ion m/z values by other mass deviations, the above procedure were repeated at least 70 times. Finally, the real FDR was calculated as the mean of the FDPs. More details are given in Supplementary information, sections 3.

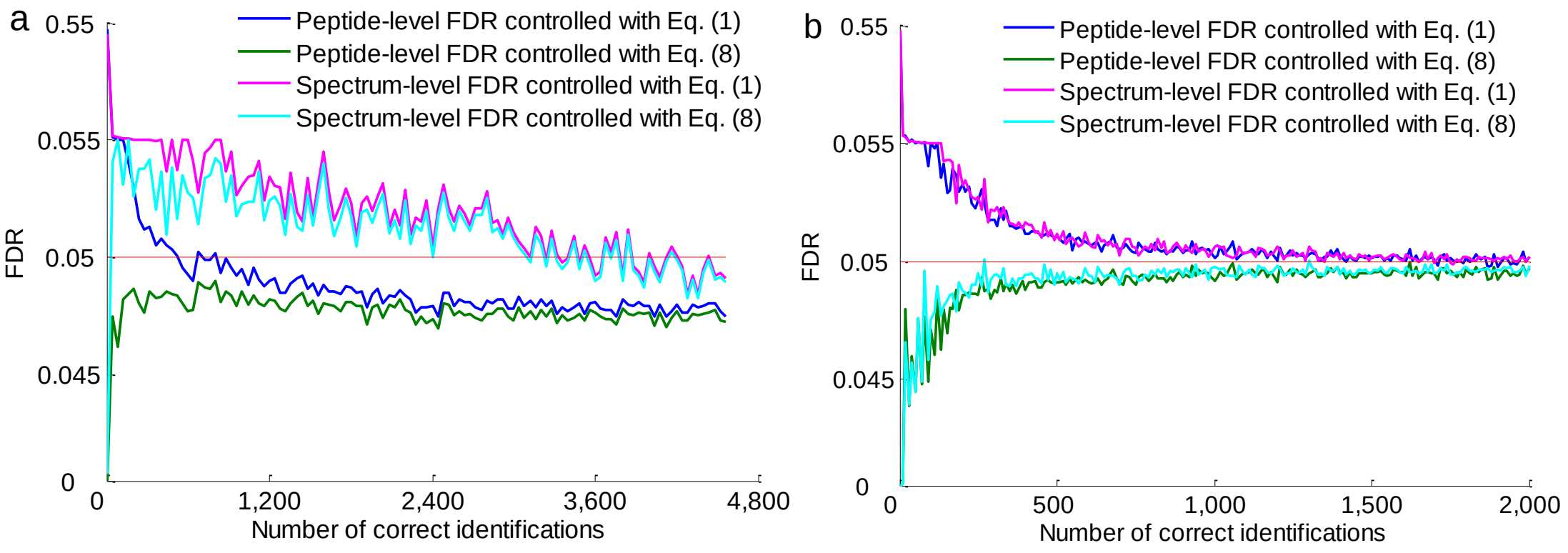

**Figure 2.** The peptide-level FDR was controlled by Eq. (8) but was not controlled by Eq. (1). The spectrum-level FDR was not controlled by both equations. The specified FDR control level is 0.05. When calculating the peptide-level FDR, different modified forms of the same peptide sequence were treated as the same peptide. (a) The Worm data set. (b) The Human data set. For the Human data set, because the correlations between the incorrect identifications are very weak, the difference between the spectrum-level FDR and the peptide-level FDR is very small. Note that the y-axis has a broken scale.

As shown by Fig. 2, the peptide-level FDR was controlled by Eq. (8) but was not controlled by Eq. (1). When the number of correct identifications is small, the peptide-level FDR controlled with Eq. (1) can be as large as 50%, far higher than the FDR control level. When the number of correct identifications is large, there is only a slight difference between the FDRs controlled with Eq. (1) and with Eq. (8).

Also shown by Fig. 2, the filtering criterion based on Eq. (8) failed to control the spectrum-level FDR. Whether the spectrum-level FDR exceeds the FDR control level depends on the data. In our experiments, the spectrum-level FDR for the Worm data set significantly exceeded the given FDR level but was almost controlled for the Human data set. The reason is that the incorrect identifications in the Worm data set were more correlated than those in the Human data set. On average, each incorrectly identified peptide in the Worm data set was incorrectly identified by 1.837 spectra while this number was only 1.051 for the Human data.

## 4 DISCUSSION

We established a theoretical foundation of the TDS strategy for FDR control in proteomics, and provided some theoretical explanations to the effectiveness and success of this strategy in practice. The difference between Eq. (8) and Eq. (1) vanishes when the data set analyzed is very large. However, for small data sets (e.g. <100 identifications), Eq. (1) may be overly optimistic sometimes. In such cases, we recommend using Eq. (8) and a higher FDR control level (e.g. 5% or 10% instead of 1%). Recently, a method named transferred FDR has been proposed for accurately estimating the FDR of a small number of protein modification identifications (Fu and Qian, 2014). However, a general approach to accurate and stable FDR estimation for small data sets of peptide identifications is still lacking.

FDR control at the spectrum level is much more difficult than at the peptide level, just as the general FDR control problem is much more difficult when the test statistics are correlated (Benjamini and Yekutieli, 2001). It is unclear whether there exists a procedure that can always control the spectrum-level FDR. With the concatenated TDS and the Eq. (8)-based filtering criterion, whether the spectrum-level FDR can be controlled in a specific experiment depends on the strength of correlations between incorrect identifications.

Our proofs about the concatenated TDS are based on the Equal Chance Assumption, but the decoy database can also be constructed in a way such that the probability of an incorrect identification being a decoy identification is $r$ times of the probability of it being a target identification (Elias and Gygi, 2007). In this case, it can be proved that

$$N_{dec}(x)/[rN_{tar}(x)] \tag{26}$$

is a conservative estimate of the FDR and the filtering criterion based on

$$[N_{dec}(x)+1]/[rN_{tar}(x)] \tag{27}$$

controls the peptide-level FDR (see Supplementary information, sections 1 for details.). Eq. (1) and Eq. (8) can be regarded as the special cases of Eq. (26) and Eq. (27), respectively, where $r = 1$.

It may seem striking that the FDR can be rigorously controlled by the concatenated TDS, which is based on a virtual coin toss rather than $p$-value. Actually, there are some deep connections between the concatenated TDS and the FDR control procedure by Storey (Storey, et al., 2004). The martingale property is the key for FDR control for both methods. The concatenated TDS can also be interpreted in a Bayesian framework (Granholm, et al., 2013;

Granholm, et al., 2011; Kall, et al., 2008). By regarding the score distribution of decoy identifications as an estimate of the null distribution, the empirical $p$-value can be calculated and the concatenated TDS can be viewed as an adaptive linear step-up procedure for FDR control (Angel, et al., 2006). (See Supplementary information, sections 2 for details.)

The separate TDS is also used for estimating and controlling the FDR of peptide identifications. But it is still unclear whether the estimate is conservative or optimistic, and whether the FDR can be rigorously controlled by the separate TDS. These problems are to be further explored.

## ACKNOWLEDGEMENTS

# A theoretical foundation of the target-decoy search strategy for false discovery rate control in proteomics

## *Supplementary Information*

## Section 1 | Proofs of Theorem 1 and Theorem 2

We first describe our model and our basic assumptions. Next we formalize Theorem 1 and Theorem 2 with the model and prove them. The symbols used in this section are independent of those used elsewhere.

## Modeling

### ➢ Formalization

Assume that we obtain $n$ identifications by the concatenated TDS and the score function is defined such that larger scores are better. Let $Id_1, Id_2, \cdots, Id_n$ be the sorted identifications and $S_1 \geq S_2 \geq \cdots \geq S_n$ be their scores. $X_i$ is defined as follows. If $Id_i$ is correct, $X_i = 0$; if $Id_i$ is an incorrect target identification, $X_i = 1$; otherwise, $Id_i$ is an incorrect decoy identification, and $X_i = -1$.

### ➢ Basic assumptions

One of the basic assumptions of the concatenated TDS is the Equal Chance Assumption. But the decoy database can also be constructed in a way such that the probability for an incorrect identification to be a decoy identification is $r$ times of the probability for it to be a target identification. In formal words, $P(X_i = -1) = rP(X_i = 1)$. The other assumption used in our proofs is the Independence Assumption. When calculating the peptide-level FDR, any two incorrectly identified spectra are matched to different peptides, and whether one spectrum is matched to the target sequences is independent of the other.

Even though the Equal Chance Assumption and the Independence Assumption hold, $S_1, S_2, \cdots, S_n, X_1, X_2, \cdots, X_n$ can be correlated. To simplify our theorems and proofs, we will first assume that $S_1, S_2, \cdots, S_n, X_1, X_2, \cdots, X_n$ are independent and then investigate the case that $S_1, S_2, \cdots, S_n, X_1, X_2, \cdots, X_n$ are correlated. A complex model is discussed in the next section for readers who are familiar with the concatenated TDS strategy and proteomics.

### ➢ Remarks

*Remark 1.* Consider identifying $n$ spectra by the concatenated TDS. When calculating the peptide-level FDR, only the best-scoring identification for each peptide is kept.

Then the number of kept identifications is a random variable related to the scores and the identified peptides. We first regard this number as a constant in our proofs and then extend our conclusions to the case that it is a random variable. Therefore, we also use $Id_1, Id_2, \cdots, Id_n$ to represent the kept identifications and $S_1 \geq S_2 \geq \cdots \geq S_n$ to represent their scores in our proofs.

*Remark 2.* Random variables $S_1, S_2, \cdots, S_n, X_1, X_2, \cdots, X_n$ are not independent of each other. There are four reasons. First, a correct identification tends to be assigned a high score. Second, when calculating the spectrum-level FDR, multiple identifications cannot be considered independent evidence of the presence of the peptide in the sample. Even for incorrect identifications, the $p$-values of the identifications of the same peptide are correlated (1, 2). Third, when calculating the peptide-level FDR, for any two different identifications of which the corresponding spectra are generated by the same peptide, at most only one of them can be correct. Last, if the size of the database is very small, the incorrectly identified spectra cannot be regarded as matched to the sequences independently when calculating the peptide-level FDR. For example, if there are only 100 peptides from the target database, at most 100 identifications can be target identifications. Fortunately, this effect is negligible when the size of database is not too small.

Though $S_1, S_2, \cdots, S_n, X_1, X_2, \cdots, X_n$ are not independent, the probability for an incorrect identification to be a target identification is usually independent of the scores of identifications and independent of which of other identifications are correct. Moreover, when calculating the peptide-level FDR, it is also independent of which of other incorrect identifications are target identifications. This is enough for our proofs as we will see. In our proofs, we will first assume that $S_1, S_2, \cdots, S_n, X_1, X_2, \cdots, X_n$ are independent and then show that the correlation between $S_1, S_2, \cdots, S_n, X_1, X_2, \cdots, X_n$ does not undermine our theorems.

## ➢ Summary

In summary, we assume that $S_1, S_2, \cdots, S_n, X_1, X_2, \cdots, X_n$ are independent random variables satisfying $S_1 \geq S_2 \geq \cdots \geq S_n$ and $P(X_i = -1) = rP(X_i = 1)$.

## Formalization of Theorem 1

Consider the identifications scoring equal to or better than score $t$ (here $t$ is a constant). If we estimate the FDR of these identifications with the concatenated TDS, we have the following conclusions.

1. With the model defined above, the number of target identifications is $\#\{X_i \neq -1 \wedge S_i \geq t\}$, the number of decoy identifications is $\#\{X_i = -1 \wedge S_i \geq t\}$, and the number of incorrect target identifications is $\#\{X_i = 1 \wedge S_i \geq t\}$. If $P(X_i = -1) = rP(X_i = 1)$, the estimated FDR for score $t$ should be $\#\{X_i = -1 \wedge S_i \geq t\}/(r\#\{X_i \neq -1 \wedge S_i \geq t\})$ (3). If the Equal Chance Assumption holds, we have $r = 1$ and the estimate is $\#\{X_i = -1 \wedge S_i \geq t\}/\#\{X_i \neq -1 \wedge S_i \geq t\}$. This is the estimate used in Theorem 1.

2. To avoid the case where the denominator is zero, we rewrite the estimated FDR as

$$\frac{\#\{X_i = -1 \wedge S_i \geq t\}}{r(\#\{X_i \neq -1 \wedge S_i \geq t\} \vee 1)} \tag{1}$$

   where $\#\{X_i \neq -1 \wedge S_i \geq t\} \vee 1 = max\{\#\{X_i \neq -1 \wedge S_i \geq t\}, 1\}$ (4). Then the expectation of the estimated FDR for fixed $t$ is

$$\frac{1}{r} E\left(\frac{\#\{X_i = -1 \wedge S_i \geq t\}}{\#\{X_i \neq -1 \wedge S_i \geq t\} \vee 1}\right). \tag{2}$$

3. The FDP of the target identifications scoring equal to or better than $t$ is

$$\frac{\#\{X_i = 1 \wedge S_i \geq t\}}{\#\{X_i \neq -1 \wedge S_i \geq t\} \vee 1} \tag{3}$$

   and the real FDR is

$$E\left(\frac{\#\{X_i = 1 \wedge S_i \geq t\}}{\#\{X_i \neq -1 \wedge S_i \geq t\} \vee 1}\right). \tag{4}$$

Theorem 1 says that for a given score $t$, the concatenated TDS provides a conservative estimate of the FDR. Namely, the expectation of the estimated FDR is not less than the real FDR. Thus, Theorem 1 can be formally expressed as follows.

*Theorem 1*. Let $S_1, S_2, \cdots, S_n, X_1, X_2, \cdots, X_n$ be independent random variables satisfying $S_1 \geq S_2 \geq \cdots \geq S_n$ and $P(X_i = -1) = rP(X_i = 1)$. Then, for any constant $t$,

$$E\left(\frac{\#\{X_i = 1 \wedge S_i \geq t\}}{\#\{X_i \neq -1 \wedge S_i \geq t\} \vee 1}\right) \leq \frac{1}{r} E\left(\frac{\#\{X_i = -1 \wedge S_i \geq t\}}{\#\{X_i \neq -1 \wedge S_i \geq t\} \vee 1}\right). \quad (5)$$

Other than the description of Theorem 1 given in the main text, the above formal representation considers the case where the probability for an incorrect identification to be a target identification is unequal to 0.5.

## Proof of Theorem 1

The proof of Theorem 1 is based on the following two lemmas.

- Lemma 1

*Lemma 1*. Let $m$ be a positive constant, and $X$ and $Y$ be two random variables satisfying $X + Y = m$ and $Y > 0$. Then

$$E\left(\frac{X}{Y}\right) \geq \frac{E(X)}{E(Y)}. \tag{6}$$

*Proof of Lemma 1*. Note that

$$E\left(\frac{X}{Y}\right) = E\left(\frac{m-Y}{Y}\right) = E\left(\frac{m}{Y}\right) - 1. \tag{7}$$

Because $m > 0$, function $g(Y) \equiv m/Y$ is convex. From Jensen's inequality, we have

$$E\left(\frac{m}{Y}\right) \geq \frac{m}{E(Y)}. \tag{8}$$

Therefore,

$$E\left(\frac{X}{Y}\right) = E\left(\frac{m}{Y}\right) - 1 \geq \frac{m}{E(Y)} - 1 = \frac{m - E(Y)}{E(Y)} = \frac{E(X)}{E(Y)}. \blacksquare \tag{9}$$

- Lemma 2

*Lemma 2*. Let $m$ be a nonnegative constant and $c$ and $r$ be positive constants. Nonnegative random variables $X, Y$ satisfy that $X + Y = m$ and $E(X) = rE(Y)$. Then

$$E\left(\frac{Y}{c+Y}\right) \leq \frac{1}{r} E\left(\frac{X}{c+Y}\right). \tag{10}$$

*Proof of Lemma 2*. Note that

$$E\left(\frac{Y}{c+Y}\right) - \frac{1}{r} E\left(\frac{X}{c+Y}\right) = E\left[\frac{rY - m + Y}{r(c+Y)}\right] = \frac{1+r}{r} - E\left[\frac{m + c + cr}{r(c+Y)}\right]. \tag{11}$$

Because $E(X) + E(Y) = m$ and $E(X) = rE(Y)$,

$$E(Y) = \frac{m}{1+r}. \tag{12}$$

Because $m$ is nonnegative and $c, r$ are positive, function $g(Y) \equiv (m + c + cr)/[r(c+Y)]$ is convex. From Jensen's inequality and Eq. (12), we have

$$E\left[\frac{m+(1+r)c}{r(c+Y)}\right] \geq \frac{m+(1+r)c}{r[c+E(Y)]} = \frac{m+(1+r)c}{r[c+m/(1+r)]} = \frac{1+r}{r}. \tag{13}$$

Finally, from Eq. (11) and Eq. (13), we have

$$E\left(\frac{Y}{c+Y}\right) - \frac{1}{r}E\left(\frac{X}{c+Y}\right) \leq \frac{1+r}{r} - \frac{1+r}{r} = 0. \blacksquare \tag{14}$$

## ➢ Proof of Theorem 1

Let $X_i = 0$ if $X_i$ is neither $1$ nor $-1$. Let $V(t) = \#\{X_i = 1 \wedge S_i \geq t\}$, $\hat{V}(t) = \#\{X_i = -1 \wedge S_i \geq t\}$, and $R(t) = \#\{X_i \neq -1 \wedge S_i \geq t\}$. For simplicity, we represent $V(t), \hat{V}(t), R(t)$ with $V, \hat{V}$ and $R$, respectively.

Note that we only need to prove the theorem under the condition that the values of $S_1, \cdots, S_n$ and the nonzero elements of $X_1, \cdots, X_n$ are fixed. Let $A_i$ be the event that $S_1, \cdots, S_n$ take some specific values and some specific elements of $X_1, \cdots, X_n$ are zero. If for any $A_i$,

$$E\left(\frac{\hat{V}}{R \vee 1} | A_i\right) \leq \frac{1}{r} E\left(\frac{V}{R \vee 1} | A_i\right), \tag{15}$$

from the law of total expectation we have

$$\begin{aligned} E\left(\frac{\hat{V}}{R \vee 1}\right) &= \sum_{A_i} E\left(\frac{\hat{V}}{R \vee 1} | A_i\right) P(A_i) \\ &\leq \sum_{A_i} \frac{1}{r} E\left(\frac{V}{R \vee 1} | A_i\right) P(A_i) = \frac{1}{r} E\left(\frac{V}{R \vee 1}\right). \end{aligned} \tag{16}$$

Therefore, we can assume that $S_1, \cdots, S_n$ are constants and the nonzero elements of $X_1, \cdots, X_n$ are fixed. Then for fixed $t$, $\#\{X_i = 0 \wedge S_i \geq t\}$ and $\#\{X_i \neq 0 \wedge S_i \geq t\}$ are also fixed and can be regarded as constants. Let $c = \#\{X_i = 0 \wedge S_i \geq t\}$ and $m = \#\{X_i \neq 0 \wedge S_i \geq t\}$. Then $V + c = R$ and $V + \hat{V} = m$.

Since for any $i$, $P(X_i = -1) = rP(X_i = 1)$, we have

$$E(\hat{V}) = rE(V). \tag{17}$$

From Lemma 1, we have

$$E\left(\frac{\hat{V}}{V}|V > 0\right) \geq \frac{E(\hat{V}|V > 0)}{E(V|V > 0)}. \tag{18}$$

Because $V$ can only take integer values, $V \geq 1$ if $V > 0$ and hence $E(V|V > 0) \geq 1$. Then from Eq. (18), we get

$$E\left(\frac{\hat{V}}{V \vee 1}\right) \geq E\left(\frac{\hat{V}}{V}|V > 0\right)P(V > 0) + E(\hat{V}|V = 0)P(V = 0)$$

$$\geq \frac{E(\hat{V}|V > 0)P(V > 0) + E(\hat{V}|V = 0)P(V = 0)}{E(V|V > 0)} = \frac{E(\hat{V})}{E(V|V > 0)}. \tag{19}$$

From Eq. (19) and Eq. (17), we have

$$\frac{1}{r}E\left(\frac{\hat{V}}{V \vee 1}\right) \geq \frac{E(\hat{V})}{rE(V|V > 0)} = \frac{E(V)}{E(V|V > 0)}. \tag{20}$$

Note that

$$\frac{E(V)}{E(V|V > 0)} = P(V > 0) = E\left(\frac{V}{V \vee 1}\right). \tag{21}$$

Hence, from Eq. (20) and Eq. (21), we have

$$E\left(\frac{V}{V \vee 1}\right) \leq \frac{1}{r}E\left(\frac{\hat{V}}{V \vee 1}\right). \tag{22}$$

Then

$$E\left[\frac{V}{(V + c) \vee 1}|c = 0\right] \leq \frac{1}{r}E\left[\frac{\hat{V}}{(V + c) \vee 1}|c = 0\right]. \tag{23}$$

If $c > 0$, because $V + \hat{V} = m$ and $E(\hat{V}) = rE(V)$, from Lemma 2 we have

$$E\left(\frac{V}{V+c}|c > 0\right) \leq \frac{1}{r}E\left(\frac{\hat{V}}{V+c}|c > 0\right). \quad (24)$$

Because $V + c = R$, from Eq. (23) and Eq. (24), we have

$$E\left(\frac{V}{R \vee 1}\right) \leq \frac{1}{r}E\left(\frac{\hat{V}}{R \vee 1}\right). \blacksquare \quad (25)$$

## ➢ Remarks

*Remark 1.* The independence of $S_1, S_2, \cdots, S_n, X_1, X_2, \cdots, X_n$ is not necessary. The assumption used in the proof is that no matter what values $S_1, S_2, \cdots, S_n$ take and which of $X_1, X_2, \cdots, X_n$ are zero, for any nonzero element $X_i$, $P(X_i = -1) = rP(X_i = 1)$ always holds. As previously described, this assumption is usually satisfied.

*Remark 2.* If $\#\{X_i = -1 \wedge S_i \geq t\}/[r(\#\{X_i \neq -1 \wedge S_i \geq t\} \vee 1)] > 1$, the FDR estimated by TDS will be larger than 1. In this case, we recommend setting the estimated FDR as 1, because the real FDR is always no more than 1. But this rarely happens. If $r = 1$, $\#\{X_i = -1 \wedge S_i \geq t\}/[r(\#\{X_i \neq -1 \wedge S_i \geq t\} \vee 1)] > 1$ means that the decoy identifications passing some threshold are more than the target identifications. This could hardly happen in reality.

*Remark 3.* In calculating the peptide-level FDR, the number of kept identifications is a random variable. Similar to the discussion of $S_1, S_2, \cdots, S_n, X_1, X_2, \cdots, X_n$, if Theorem 1 holds for any constant number of identifications, it also holds when the number is a random variable.

## Formalization of Theorem 2

If we control the FDR with the filtering criterion defined in Theorem 2, we have the following conclusions.

1. Assume that $S_k$ is chosen as the score threshold. Then the identifications passing the score threshold are $Id_1, Id_2, \cdots, Id_k$. The number of decoy identifications is $\#\{X_i = -1 \wedge i \leq k\}$, the number of target identifications is $\#\{X_i \neq -1 \wedge i \leq k\}$, and the estimated FDR is $[\#\{X_i = -1 \wedge i \leq k\} + 1]/\#\{X_i \neq -1 \wedge i \leq k\}$. To avoid the case where $\#\{X_i \neq -1 \wedge i \leq k\} = 0$ and $r \neq 1$, we rewrite the estimated FDR as
$$\frac{1}{r} \times \frac{\#\{X_i = -1 \wedge i \leq k\} + 1}{\#\{X_i \neq -1 \wedge i \leq k\} \vee 1}. \quad (26)$$
2. Let
$$K = max\left\{k | \frac{1}{r} \times \frac{\#\{X_i = -1 \wedge i \leq k\} + 1}{\#\{X_i \neq -1 \wedge i \leq k\} \vee 1} \leq \alpha\right\}. \quad (27)$$
Then $S_k$ is the lowest *x* at which the estimated FDR is no more than $\alpha$. Thus, the final score threshold is $S_k$ and $Id_1, Id_2, \cdots, Id_K$ will pass the threshold.
3. Note that the FDR of $Id_1, Id_2, \cdots, Id_K$ is
$$E\left(\frac{\#\{X_i = 1 \wedge i \leq K\}}{\#\{X_i \neq -1 \wedge i \leq K\} \vee 1}\right). \quad (28)$$

Theorem 2 says that for a given score $\alpha$, retaining $Id_1, Id_2, \cdots, Id_K$ can control the FDR; that is, the FDR of $Id_1, Id_2, \cdots, Id_K$ is not more than $\alpha$. Then Theorem 2 can be formally expressed as follows.

*Theorem 2.* Let $X_1, X_2, \cdots, X_n$ be independent random variables such that $P(X_i = -1) = rP(X_i = 1)$. For any $\alpha$, define

$$K = max\left\{k | \frac{1}{r} \times \frac{\#\{X_i = -1 \wedge i \leq k\} + 1}{\#\{X_i \neq -1 \wedge i \leq k\} \vee 1} \leq \alpha\right\}. \quad (29)$$

If there is no such $k$, let $K = 0$. Then

$$E\left(\frac{\#\{X_i = 1 \wedge i \leq K\}}{\#\{X_i \neq -1 \wedge i \leq K\} \vee 1}\right) < \alpha. \tag{30}$$

Other than the description of Theorem 2 given in the main text, the above formal representation considers the case where the probability for an incorrect identification to be a target identification is unequal to 0.5.

## Proof of Theorem 2

The proof of Theorem 2 is based on the following two lemmas.

- Lemma 3

*Lemma 3.* Let $X_1, X_2, \cdots, X_m$ be independent random variables such that $P(X_i = -1) = r/(r+1), P(X_i = 1) = 1/(r+1)$. Let $L$ be the maximum number such that $X_1, X_2, \cdots, X_L$ are not $-1$. Then $E(L) < 1/r$.

*Proof of Lemma 3.* For any $i < m$, the probability that $X_1 = X_2 = \cdots = X_i = 1$ and $X_{i+1} = -1$ is $1/(r+1)^i \times r/(r+1)$. The probability that $X_1 = X_2 = \cdots = X_m = 1$ is $1/(r+1)^m$. Then, we have

$$E(L) = \sum_{i=0}^{m-1} i \left(\frac{1}{r+1}\right)^i \frac{r}{r+1} + m \left(\frac{1}{r+1}\right)^m. \quad (31)$$

Since

$$\sum_{i=0}^{m-1} i \left(\frac{1}{r+1}\right)^i = \frac{r+1}{r^2}\left[1 - \frac{mr+1}{(r+1)^m}\right], \quad (32)$$

we have

$$E(L) < \frac{r+1}{r^2}\left[1 - \frac{mr}{(r+1)^m}\right]\frac{r}{r+1} + m\left(\frac{1}{r+1}\right)^m = \frac{1}{r}. \quad (33)$$

The meaning of Lemma 3 is as follows. If the probability for an incorrect identification to be a target identification is $1/(r+1)$, the expected number of incorrect target identifications scoring better than the best-scoring decoy identification is smaller than $1/r$ (when there is no decoy identification, it is the number of all incorrect target identifications).

- Lemma 4

*Lemma 4.* Let $X_1, X_2, \cdots, X_m$ be independent random variables such that $P(X_i =$

$-1) = p, P(X_i = 1) = 1 - p$. Let $L$ be the maximum number such that $X_1, X_2, \cdots, X_L$ are not $-1$. Then for any nonnegative constants $a$ and $b$ satisfying $a + b = m$ or $X_{a+b+1} = -1$,

$$E\left[L \middle| \sum_{i=1}^{a+b} \mathbb{1}(X_i = 1) = a\right] = \frac{a}{b+1}. \tag{34}$$

*Proof of Lemma 4*. Because $X_1, X_2, \cdots, X_m$ are independent and identically distributed, if $a + b = m$ or $X_{a+b+1} = -1$, under the condition $\sum_{i=1}^{a+b} \mathbb{1}(X_i = 1) = a$, the probabilities for $X_1, X_2, \cdots, X_{a+b}$ to be all sequences consisting of $a$ elements equal to 1 and $b$ elements equal to $-1$ are the same. Then the $b$ elements equal to $-1$ separate the $a$ elements equal to $1$ into $b + 1$ intervals, and for any element equal to $1$, it falls into each interval with the same probability. Then the probability is $1/(b + 1)$, and the expected number of elements falling into each interval is $a/(b + 1)$. $L$ is the number of elements equal to $1$ which fall into the interval before the first $-1$, then the expectation of $L$ is also $a/(b + 1)$.

The meaning of Lemma 4 is as follows. When $a$ incorrect target identifications and $b$ decoy identifications pass the score threshold, the expected number of incorrect target identifications scoring better than the best-scoring decoy identification is $a/(b + 1)$.

## ➢ Proof of Theorem 2

Let $X_i = 0$ if $X_i$ is neither $1$ nor $-1$. Similar to the proof of Theorem 1, we only need to prove the theorem under the condition that the nonzero elements in $X_1, X_2, \cdots, X_n$ are fixed.

Let $V = \#\{X_i = 1 \wedge i \le K\}$, $\hat{V} = \#\{X_i = -1 \wedge i \le K\}$, and $R = \#\{X_i \ne -1 \wedge i \le K\}$. Consider the nonzero elements in $X_1, X_2, \cdots, X_n$. If $X_i$ is a nonzero element, then $P(X_i = -1) = r/(1 + r), P(X_i = 1) = 1/(1 + r)$. At the same time, because $X_1, X_2, \cdots, X_n$ are independent, we have that the nonzero elements in them are also independent. Assume $m$ is the total number of nonzero elements. Let $L$ be the number of elements equal to 1 before the first element equal to $-1$. In formal words, $L = \#\{X_i = 1 \wedge i \leq j\}$ where $j$ is the maximum number such that $X_1, X_2, \cdots, X_j$ are not $-1$.

If $K = 0$, we have $V = 0, R = 0$ and $\hat{V} = 0$. In this case,

$$E(L|K = 0) \geq 0. \tag{35}$$

If $K = n$, then $V + \hat{V} = m$. Therefore, by applying Lemma 4 (in the case $a + b = m$) to the $m$ nonzero elements in $X_1, X_2, \cdots, X_n$, we have

$$E\big(L\big|V = a, \hat{V} = b, K = n\big) = \frac{a}{b+1}. \tag{36}$$

If $0 < K < n$, $X_{K+1} = -1$. Otherwise, $X_{K+1} \neq -1$, then $\#\{X_i = -1 \wedge i \leq K + 1\} = \#\{X_i = -1 \wedge i \leq K\}$ and $\#\{X_i \neq -1 \wedge i \leq K + 1\} \vee 1 = \#\{X_i \neq -1 \wedge i \leq K\} + 1$. Therefore,

$$\frac{1}{r} \times \frac{\#\{X_i = -1 \wedge i \leq K + 1\} + 1}{\#\{X_i \neq -1 \wedge i \leq K + 1\} \vee 1} = \frac{1}{r} \times \frac{\#\{X_i = -1 \wedge i \leq K\} + 1}{\#\{X_i \neq -1 \wedge i \leq K\} + 1}$$

$$\leq \frac{1}{r} \times \frac{\#\{X_i = -1 \wedge i \leq K\} + 1}{\#\{X_i \neq -1 \wedge i \leq K\} \vee 1} \leq \alpha. \tag{37}$$

This contradicts with Eq. (29). Then $X_{K+1} = -1$. Note that $X_{K+1}$ is the $\big(V + \hat{V} + 1\big)$-th nonzero element in $X_1, X_2, \cdots, X_n$. By applying Lemma 4 (in the case $X_{a+b+1} = -1$) to the $m$ nonzero elements in $X_1, X_2, \cdots, X_n$, we have

$$E\big(L\big|V = a, \hat{V} = b, 0 < K < n\big) = \frac{a}{b+1}. \tag{38}$$

From Eq. (35), Eq. (36) and Eq. (38), we have that no matter what integer values $K$ takes,

$$E\big(L\big|V = a, \hat{V} = b\big) \geq \frac{a}{b+1}. \tag{39}$$

Therefore,

$$E(L) = \sum_{a,b} E\big(L \big| V = a, \hat{V} = b\big) P\big(V = a, \hat{V} = b\big)$$

$$\geq \sum_{a,b} \frac{a}{b+1} P\big(V = a, \hat{V} = b\big) = E\left(\frac{V}{\hat{V}+1}\right). \quad (40)$$

By applying Lemma 3 to the $m$ nonzero elements in $X_1, X_2, \cdots, X_n$, we have $E(L) < 1/r$. Then

$$E\left(\frac{V}{\hat{V}+1}\right) \leq E(L) < \frac{1}{r}. \quad (41)$$

Since $\hat{V} = \#\{X_i = -1 \wedge i \leq K\}$ and $R = \#\{X_i \neq -1 \wedge i \leq K\}$, from Eq. (29) we have

$$\frac{1}{r} \times \frac{\hat{V}+1}{R \vee 1} \leq \alpha. \quad (42)$$

Namely, if $R > 0$,

$$R \geq \frac{\big(\hat{V}+1\big)}{r\alpha}, \quad (43)$$

then

$$\frac{V}{R \vee 1} \leq \frac{r\alpha V}{\hat{V}+1}. \quad (44)$$

Because $V \leq R$, if $R = 0$, we have $V = 0$ and $V/(R \vee 1) = 0 = r\alpha V/(\hat{V}+1)$. Namely, if $R = 0$, Eq. (44) also holds. Then from Eq. (41) and Eq. (44) we have

$$E\left(\frac{V}{R \vee 1}\right) \leq E\left(\frac{r\alpha V}{\hat{V}+1}\right) < r\alpha \times \frac{1}{r} = \alpha. \blacksquare \quad (45)$$

## ➢ Remarks

*Remark 1.* Because the values of $S_1, S_2, \cdots, S_n$ do not affect the value of $K$ or the conditional probability $P(X_i = -1 | X_i \neq 0)$, we can rewrite Theorem 2 without $S_1, S_2, \cdots, S_n$.

*Remark 2.* The independence of $X_1, X_2, \cdots, X_n$ is unnecessary. Our proof of Theorem 2 only uses the independence of the nonzero elements in $X_1, X_2, \cdots, X_n$, which is equivalent to the Independence Assumption.

*Remark 3.* In calculating the peptide-level FDR, the number of kept identifications is a random variable. Similar to the discussion of Theorem 1, we only need to prove Theorem 2 for the case where this number is a constant.

## Formalization of Theorem 3

Similar to Theorem 2, Theorem 3 can be formally expressed as follows.

*Theorem 3*. Let $X_1, X_2, \cdots, X_n$ be independent random variables such that $P(X_i = -1) = rP(X_i = 1)$. For any $\alpha$, define

$$K = max\left\{k \middle| \frac{1}{r} \times \frac{\#\{X_i = -1 \wedge i \leq k\} + c}{\#\{X_i \neq -1 \wedge i \leq k\} \vee 1} \leq \alpha\right\} \tag{46}$$

where $c < 1$. If there is no such $k$, let $K = 0$. Then there are some $X_1, X_2, \cdots, X_n$ such that

$$\mathrm{E}\left(\frac{\#\{X_i = 1 \wedge i \leq K\}}{\#\{X_i \neq -1 \wedge i \leq K\} \vee 1}\right) > \alpha. \tag{47}$$

- Proof of Theorem 3

We prove Theorem 3 with a counter-example. Let $X_1, X_2, \cdots, X_m$ are 0 and $X_{m+1}, X_{m+2}, \cdots, X_n$ are either 1 or $-1$ where $r\alpha m \geq c$ and

$$r(m+n)\alpha < 1 - \frac{1}{(r+1)^n}. \tag{48}$$

For example, let $n = \lceil -log_{r+1}(0.4 - 0.4c) \rceil, m = \lceil 2cn/(1-c) \rceil$ and $\alpha = c/rm$, it is not difficult to verify that $r\alpha m \geq c$ and Eq. (48) are satisfied.

Let $T$ be the number of 1 before the first $-1$. If there is no $-1$ in $X_{m+1}, X_{m+2}, \cdots, X_n$, let $T$ be the number of 1 in $X_{m+1}, X_{m+2}, \cdots, X_n$. Because

$$\frac{1}{r} \times \frac{\#\{X_i = -1 \wedge i \leq m + T\} + c}{\#\{X_i \neq -1 \wedge i \leq m + T\} \vee 1} = \frac{c}{r(m+T)} \leq \frac{c}{rm} \leq \alpha, \tag{49}$$

we have $K \geq T + m$ and $\#\{X_i = 1 \wedge i \leq K\} \geq T$. Then

$$\mathrm{E}\left(\frac{\#\{X_i = 1 \wedge i \leq K\}}{\#\{X_i \neq -1 \wedge i \leq K\} \vee 1}\right) = \mathrm{E}\left(\frac{\#\{X_i = 1 \wedge i \leq K\}}{m + \#\{X_i = 1 \wedge i \leq K\}}\right) \geq \mathrm{E}\left(\frac{T}{m+T}\right). \tag{50}$$

From Eq. (31), we have

$$E(T) = \sum_{i=0}^{n-1} i \left(\frac{1}{r+1}\right)^{i} \frac{r}{r+1} + n \left(\frac{1}{r+1}\right)^{n} = \frac{1}{r} - \frac{1}{r}\frac{1}{(r+1)^{n}}. \qquad (51)$$

Because $T \leq n$, we have

$$\mathrm{E}\left(\frac{T}{m+T}\right) \geq \mathrm{E}\left(\frac{T}{m+n}\right) = \frac{\mathrm{E}(T)}{m+n} = \frac{1}{r(m+n)} \times \left[1 - \frac{1}{(r+1)^{n}}\right]. \qquad (52)$$

From Eq. (48), we have

$$\mathrm{E}\left(\frac{T}{m+T}\right) \geq \frac{1}{r(m+n)} \times \left[1 - \frac{1}{(r+1)^{n}}\right] > \alpha. \blacksquare \qquad (53)$$

## Section 2 | Connections between the concatenated TDS and multiple testing methods

➢ The adaptive linear step-up procedure

By introducing the empirical $p$-value, the procedure of controlling the FDR with the concatenated TDS can be viewed as an adaptive linear step-up procedure (5). The procedure first estimates the number of incorrect identifications, and then uses this estimate to improve the BH procedure. Assume that the FDR control level is $\alpha$. Let $N_{dec}$ be the total number of decoy identifications, $N_{inc}$ be the total number of incorrect target identifications, and $N_{tar}$ be the total number of target identifications. The $p$-value of score $x$ can be estimated as

$$\widehat{p_x} = \frac{N_{dec}(x) + 1}{N_{dec} + 1}, \tag{54}$$

where $\widehat{p_x}$ is the empirical $p$-value (6). The BH procedure chooses the minimum $x$ satisfying

$$\widehat{p_x} \leq \frac{N_{tar}(x)\alpha}{N_{tar}} \tag{55}$$

as the score threshold ($N_{tar}(x)$ is $i$, $\widehat{p_x}$ is $p_{(i)}$, $N_{tar}$ is $m$ and $\alpha$ is $q^*$) (7). The filtering criterion defined in Theorem 2 chooses

$$t = min\left\{x| \frac{N_{dec}(x) + 1}{N_{tar}(x)} \leq \alpha\right\} \tag{56}$$

as the score threshold. From Eq. (54) and Eq. (56), we have that $t$ is the minimum $x$ satisfying

$$\widehat{P(x)} \leq \frac{N_{tar}(x)\alpha}{N_{dec} + 1}. \tag{57}$$

Between Eq. (55) and Eq. (57), the only difference is the denominator, which can be viewed as the estimated number of incorrect target identifications. For the BH procedure, the estimate is $N_{tar}$, but $N_{tar}$ is usually much more larger than $N_{inc}$ because there are usually some correct target identifications. While for the filtering criterion defined in Theorem 2, the estimate is $N_{dec} + 1$. Because the number of incorrect target identifications is approximately equal to the number of decoy identifications, $N_{dec} + 1$ is a reasonable estimate for $N_{inc}$. As $N_{dec} + 1$ is usually less than $N_{tar}$, the score threshold chosen by the filtering criterion defined in Theorem 2 is usually smaller than that chosen by the BH procedure. Namely, with the filtering criterion defined in Theorem 2, one can obtain more identifications than with the BH procedure.

As shown in Eq. (57), the filtering criterion defined in Theorem 2 estimates the number of incorrect target identifications as $N_{dec} + 1$. In other words, the proportion of incorrect target identifications is estimated as $(N_{dec} + 1)/N_{tar}$. Therefore, just as the separate TDS and some general FDR control methods (4, 5, 8), the proportion of incorrect target identifications is also used by the concatenated TDS to enhance the number of retained identifications at a given FDR level.

- ## The martingale property

Both the procedure introduced by Benjamini and Hochberg (the BH procedure) and the procedure introduced by Storey define a martingale (9). In fact, the procedure of controlling the FDR with the concatenated TDS also defines a martingale and the martingale property has been used in our proof of Theorem 2 (the proof of Eq. (36) and Eq. (38) with lemma 4).

Let $T_i$ be the number of incorrect target identifications scoring better than the $i$-th best-scoring decoy identification. If there are less than $i$ decoy identifications in total, $T_i$ is the number of all incorrect target identifications. For example, $T_1$ is the number of incorrect target identifications scoring better than the best-scoring decoy identifications. Let $D$ be the total number of decoy identifications passing the score threshold.

According to the Equal Chance Assumption and the Independence Assumption,

for any $i \leq b$, we have

$$E(T_1|D = b, T_{D+1} = a) = \frac{a}{b+1}. \quad (58)$$

Namely, the $b$ decoy identifications divide the $a$ incorrect target identifications into $b+1$ intervals, and the expected numbers of incorrect target identifications falling into all intervals are the same. For example, suppose that two incorrect target identifications and one decoy identification pass the score threshold, i.e., $a = 2$ and $b = 1$. For the two target identifications and the best-scoring decoy identification, there are three possible cases: i) two target identifications both score better than the decoy identification, ii) only one target identification scores better than the decoy identification and iii) no target identification scores better than the decoy identification. The probabilities of all these three cases are the same. Therefore, the expectation of $T_1$ is $(2+1+0)/3 = 1$, equal to $a/(b+1) = 2/2$.

Similarly, we can also prove that for any $1 < i \leq D+1$,

$$E\left(\frac{T_{i-1}}{i-1}|T_i, \cdots, T_{D+1}\right) = \frac{T_i}{i}, \quad (59)$$

because the expected numbers of incorrect target identifications falling into all the $i$ intervals are all $T_i/i$. Therefore, for fixed $D$, $T_1, T_2/2, \cdots, T_{D+1}/(D+1)$ is a martingale with time running backwards.

## Section 3 | Details of experiments

Our experiment involves two steps. The first step is to obtain reliably identified spectra (RI spectra) and unreliably identified spectra (UI spectra) with the common search process. The second step is to calculate the real FDR by repeating the search and the filtration. The database search parameters are listed in **Table 1**.

To calculate the real FDR, we need to repeat the search and the filtration many times. If all incorrect matches in the search have random scores, a spectrum with a correct match may be incorrectly identified, because some incorrect matches may score better than the correct match just by chance. Therefore, the number of correct identifications is a random variable. But the real FDR is influenced by the number of correct identifications. To show this influence, we fixed this number in calculating an FDR and then calculated FDRs for different numbers.

To calculate an FDR for a number $n$, $n$ RI spectra were sampled. For these RI spectra, the UI spectra were sampled multiple times without replacement to repeat the search. When calculating the spectrum-level FDR, the spectra were sampled in a manner such than the spectra identified as the same peptides were used simultaneously. The number of UI spectra used in a search also affected the spectrum-level FDR. For $n$ RI spectra, the UI spectra identified as $m$ different peptides were used in a search. For the worm data set, $m = min\{400, max\{100,0.5n\}\}$. For Human data set, $m = min\{n + 100, max\{100,0.2n\}\}$.

In the second step, to repeat the search and obtain enough instances of FDPs, the precursor ion m/z values of the UI spectra were shifted by different mass deviations. For the Worm data set, the precursor ion m/z values were shifted by 6, 7, 8, 9, 10, 11, 12, 13, 15 and 17 Da respectively in different repetitions. For the Human data set, the precursor ion m/z values were shifted by $-10$, $-9$, $-8$, $-7$, $-6$, 6, 7, 8, 9, 10, 11, 12, 13, 15 and 17 Da respectively in different repetitions. We avoided the mass deviations close to the masses of some common modifications, such as Methylation (14 Da) and Oxidation (16 Da). Even though the precursor ion m/z values are incorrect, some spectra may still be correctly identified because the peptides are chemically modified. To remove the correct results, we weeded out all the identifications with $E$-values less than 0.001.

For the Worm data set, each FDR is the mean of at least 70 FDPs. The number of FDPs depends on the size of RI spectra $n$. For larger $n$, more UI spectra are sampled for a search. And then the number of FDPs we can obtain is less. For the Human data set, each FDR is the mean of at least 250 FDPs. To calculate the FDR more precisely, we need more repetitions. If the FDR threshold is 1%, more repetitions are necessary to reduce the variance. So we set the FDR threshold at 5% in the second step. Compared with the Worm data set, the real FDRs of the Human data set were closer to 5%. To reveal the slight difference between the real FDR and the specified FDR control level, more repeated searches were performed for the Human data set.

For the Human data set, 4,879 of the 7,833 spectra were reliably identified in the common search process, and the remaining 2954 spectra were unreliably identified. To obtain enough incorrect identifications, except 2000 RI spectra that were randomly selected, the precursor ion m/z values of all other spectra were shifted in subsequent searches.

**Table 1|** Parameters for database search by pFind 2.8 and MaxQuant 1.3.0.5

| Item | Setting |
|---|---|
| Enzyme | Trypsin |
| Maximum missed cleavage sites | 2 |
| Precursor tolerance | $\pm$20 ppm for pFind, $\pm$20 ppm for the first search of MaxQuant and $\pm$6 ppm for the main search of MaxQuant |
| Fragment tolerance | $\pm$20 ppm |
| Fixed modifications | Carbamidomethyl (C) |
| Variable modifications | Oxidation (M) for the Worm data set, Oxidation (M) and Acetyl (protein N-term) for the Human data set |